\documentclass[usegraphicx,usenatbib]{mn2e}

\title[Galaxy Evolution at $z < 0.3$]
{Evolution of the Galaxy Luminosity Function at $\bmath{z < 0.3}$}

\author[J. Loveday]
{
Jon Loveday\thanks{E-mail: J.Loveday@sussex.ac.uk}\\
Astronomy Centre, University of Sussex, Falmer, Brighton, BN1 9QJ
}

\begin{document}

\maketitle

\begin{abstract}
We measure the redshift-dependent luminosity function and 
the comoving radial density of galaxies in the Sloan Digital Sky Survey 
Data Release 1 (SDSS DR1).
Both measurements indicate that the apparent number density of bright 
galaxies increases by 
a factor $\approx 3$ as redshift increases from $z=0$ to $z=0.3$.
This result is robust to the assumed cosmology, to the details
of the $K$-correction and to direction on the sky.
These observations are most naturally explained by significant evolution 
in the luminosity and/or number density of galaxies at redshifts $z < 0.3$.
Such evolution is also consistent with the steep number-magnitude counts
seen in the APM Galaxy Survey, without the need to invoke a local underdensity
in the galaxy distribution or magnitude scale errors.
\end{abstract}


\section{Introduction}

Measurements of the galaxy luminosity function (LF) and its evolution
provide important constraints on theories of galaxy formation and evolution.
It is currently believed that galaxies formed hierarchically
from the merger of sub-clumps, with the peak of star formation rate
occurring around redshifts $z \approx 2$--4, eg. \citet{clbf2000}.
Since then, galaxies are thought to have evolved mostly passively
as their stellar populations age, with occasional activity
triggered by interactions with other galaxies.

Significant evolution in the LF has been measured since redshift
$z \sim 1$ \citep[eg.][]{lilly95,ecbhg96,wolf2003},
but most existing galaxy samples have been too small to directly constrain 
evolution at more recent epochs.
By combining three different redshift surveys, \citet{eales1993}
was able to demonstrate that the amplitude of the LF increases by a
factor $\approx 3$ in the redshift range $0 < z < 0.4$.
He also found evidence for an increase in the amplitude of the LF to
$z \approx 0.15$--0.2, although the strength of this low-redshift
evolution was poorly constrained.

The Sloan Digital Sky Survey \citep[SDSS,][]{york2000}, provides an ideal
sample with which to measure galaxy evolution at low redshifts.
A recent determination of the LF at redshift $z=0.1$ \citep{blan2003L},
allowing for galaxy evolution, showed that the typical $r$-band 
galaxy luminosity brightens by $\approx 1.6$ mag per unit redshift.
Here we further investigate evolution in the $r$-band LF of SDSS galaxies.
We describe the data sample in Section~\ref{sec:data} and estimate the
LF in four redshift slices in  Section~\ref{sec:lf}.
The effects of galaxy evolution are investigated in an alternative way in
Section~\ref{sec:dens}, where we estimate the radial density of galaxies.
Comparisons with observed number-magnitude counts are made in
Section~\ref{sec:counts} and we conclude in Section~\ref{sec:concs}.

\section{Data Sample} \label{sec:data}

We use galaxies from the Sloan Digital Sky Survey
Data Release 1 \citep[DR1,][]{abazajian2003}.
The SDSS is performing five band CCD imaging over an area $\sim 10,000$
square degrees (\citealt{fuku96},
\citealt{gunn98}, \citealt{hfsg2001},
\citealt{smith2002}, \citealt{pier2003}).  
A multifibre spectrograph is measuring spectra and redshifts
for a subset of sources detected in the imaging data;
here we consider only galaxies in the main flux-limited sample
\citep{strauss2002}.
A technical summary of the survey is given in \citet{york2000} and
a description of SDSS photometric and spectroscopic parameters
may be found in \citet{stough2002}.

The magnitude limit of the main galaxy survey has been set at an
extinction-corrected \citep{sfd98} $r$-band Petrosian magnitude
$r < 17.77$.
This magnitude limit was chosen as test year data demonstrated that it
corresponds closely to the desired target density of 90 objects per
square degree, see \citet{strauss2002} for full details of the target 
selection algorithm.
However, a small amount of early data included in DR1 was taken when 
the magnitude limit was set at $r < 17.6$, and so we adopt this
brighter magnitude limit here.
We thus select galaxies from the DR1 catalogue with extinction-corrected 
$r$-band Petrosian magnitude $r < 17.6$ and with one or more of the 
\verb|TARGET_GALAXY|, \verb|TARGET_GALAXY_BIG| or 
\verb|TARGET_GALAXY_BRIGHT_CORE| bits set in the {\tt primTarget} bit mask;
see \citet{stough2002} for a description of these target classifications.

These selection criteria yield a sample of 162,989 target galaxies over
2099 square degrees.
Of these target galaxies, 91,921 have had a spectrum observed and
91,611 have a redshift measured with a confidence of 80\% or higher.
The sampling rate for our sample, defined as the number of galaxies
with reliable redshifts divided by the number of target galaxies,
is thus $f = 0.562$.
This low sampling rate is simply due to the fact that SDSS spectroscopic
observations lag imaging observations; the effective spectroscopic area
is 1360 square degrees.
The fraction of targets observed spectroscopically is further reduced by
limitations on the placement of spectroscopic fibres \citep{blmyzl2003}.
The fraction of spectra yielding reliable redshifts is more than 99.6\%,
with no discernible dependence on apparent magnitude.
Whilst we naturally cannot demonstrate that spectroscopic success rate is
independent of redshift, the fraction of galaxies without reliable
redshifts is only 0.4\%, and so spectroscopic incompleteness will have
a negligible effect on estimates of galaxy evolution.

Heliocentric velocities are converted to the Galactocentric frame
using $v_{\rm Gal} = v_{\rm Helio} + 220 \sin(l) \cos(b)$.
Unless otherwise stated, we assume a Hubble constant of $H_0 = 100$
km/s/Mpc and an $\Omega_M = 0.3, \Omega_\Lambda = 0.7$ cosmology in
calculating distances, comoving volumes and luminosities.

When estimating intrinsic galaxy luminosities, it is necessary to correct
for the fact that a fixed observed passband corresponds to a different
range of wavelengths in the rest frames of galaxies at different redshifts,
the so-called $K$-correction.
The $K$-correction depends on the passband used, the redshift of the galaxy
and its spectral energy distribution (SED).
Here we use \verb|kcorrect v3_1b| \citep{blan2003K} in order to estimate
and apply these corrections.
Briefly, this code estimates an SED for each galaxy by finding the
non-negative, linear combination of three template spectra that gives 
the best-fit to the five SDSS magnitudes of that galaxy.
Rather than estimating luminosities in the rest-frame of each galaxy, 
we use \verb|kcorrect| to estimate luminosities in a passband blue-shifted
by $z=0.1$.
Following \citet{blan2003K} we denote the $r$-band in this frame as $^{0.1}r$.
The advantage of this choice of restframe is that galaxies at redshift
$z=0.1$ (close to the mean for the SDSS main galaxy sample) have 
$K$-corrections independent of galaxy type, and these corrections
are on average smaller in amplitude than $K$-corrections at redshift zero.

\section{Galaxy luminosity function} \label{sec:lf}

We estimate the LF using the \citet[STY]{sty79}
parametric maximum likelihood method and the stepwise maximum likelihood (SWML)
method of \citet[EEP]{eep88}.
These estimators
are unbiased by density inhomogeneities and have well-defined error properties.
Both methods assume that $\phi(L)$ has a universal form, {\it i.e.} the
number density of galaxies is separable into a function of luminosity times a
function of position: $n(L, {\bf x}) = \phi(L) \rho({\bf x})$. 
Using these estimators, the shape of $\phi(L)$ is determined independently
of its normalization.

\subsection{Shape}

The probability of seeing a galaxy of luminosity $L_i$ at redshift
$z_i$ in a flux-limited catalogue is given by
\begin{equation}
 p_i \propto \phi(L_i) \left/ \int_{L_{\rm min}(z_i)}^{L_{\rm max}(z_i)}
      \phi(L) dL\right., \label{eqn:STY}
\end{equation}
where $L_{\rm min}(z_i)$ and $L_{\rm max}(z_i)$ are
the minimum and maximum luminosities observable at redshift $z_i$
in a flux limited sample. In the STY method, the likelihood
${\cal L} = \prod p_i$ (where the product extends over all galaxies
in the sample) is maximized with respect to a set of parameters 
describing the function $\phi(L)$. For example, if we
assume that $\phi(L)$ is described by a \citet{schec76} function,
\begin{equation}
\phi(L) dL = \phi^* \left(\frac{L}{L^*} \right)^{\alpha} \exp
\left(-\frac{L}{L^*} \right) d\left(\frac{L}{L^*} \right),
\label{eqn:schec}
\end{equation}
we maximize the likelihood with respect to $\alpha$ and $L^*$.
Errors in the Schechter parameters are estimated by the jackknife method,
whereby we subdivide the galaxies into 20 roughly equal sub-samples
and estimate the Schechter parameters omitting each sub-sample in turn.
The variance in parameter $x$ is then given by
\begin{equation}
{\rm Var}(x) = \frac{N-1}{N} \sum_{i=1}^N (x - \bar{x})^2,
\end{equation}
where $N=20$ is the number of sub-samples and $\bar{x}$ is the mean of $x$.

For the highest redshift sub-sample considered here, $z > 0.2$,
only galaxies more luminous than $M_{^{0.1}r} \approx -21.5$
make it into the SDSS main galaxy sample.
We thus have only a range of about two magnitudes over which to fit
the Schechter function.
Consequently, the accuracy with which the Schechter parameters 
can be determined is limited, particularly the value of the
faint-end slope $\alpha$.
We therefore also estimate the luminosity function using
the SWML method of EEP in which $\phi(L)$ is parameterized
as a set of numbers $\phi_k$ in equally spaced magnitude bins. 
The likelihood ${\cal L}$ is maximized with respect to
$\phi_k$ applying constraints as described in EEP. 
Also following EEP, errors on the $\phi_k$ are estimated from the 
information matrix.

\subsection{Normalization}

We use the following estimator of the space density $\bar{n}$ of galaxies:
\begin{equation}
 f\bar{n} = \sum_{i=1}^{N_{\rm gal}} w(z_i) \left/ 
            \int_{z_{\rm min}}^{z_{\rm max}} dV S(z) w(z)\right.,
 \label{eqn:nbar}
\end{equation}
where $f$ is the sampling rate, $S(z)$ the galaxy selection function
and $w(z)$ a weighting function.
The selection function for galaxies with luminosities $L_1$ to $L_2$
is 
\begin{equation}
 S(z) = \int_{{\rm max}(L_{\rm min}(z),L_1)}^{{\rm min}(L_{\rm max}(z),L_2)}
        \phi(L) dL \left/ \int_{L_1}^{L_2} \phi(L) dL\right..
\label{eqn:selfn}
\end{equation}
Note that the integration limits in the numerator depend on the $K$-correction.
In this case, we estimate the $K$-correction using an SED created from
the mean template coefficients of all galaxies in the sample.

We adopt the weighting function
\begin{equation}
 w(z) = {1 \over [1 + 4 \pi f \bar{n} J_3(r_c) S(z)]}, 
\quad J_3(r_c) = \int_0^{r_c} r^2 \xi(r) dr
\label{eqn:weight}
\end{equation}
where $\xi(r)$ is the two point galaxy correlation function. Provided
$J_3(r_c)$ converges on a scale $r_c$ much smaller than the depth of
the survey, then the weighting scheme (eq.~\ref{eqn:weight})
minimizes the variance in the estimate of $\bar {n}$ \citep{dh82}.
Larger values of $J_3$ weight galaxies at high redshift more highly;
we adopt $4\pi J_3 \approx 32,000 h^{-3} {\rm Mpc}^3$.
This value comes from integrating the two-point galaxy correlation function
of \citet{zehavi2002}, $\xi(r) = (r/6.14)^{-1.75}$, to 
$r_c = 50 h^{-1} {\rm Mpc}$; 
at larger separations the value of $J_3$ becomes uncertain. 
However, the results are not too sensitive to the value
of $J_3$, the estimated density decreasing by 7\% if $J_3$ is halved.

When normalizing the non-parametric SWML estimate, the integrals in
(\ref{eqn:selfn}) are replaced by sums over magnitude bins, with
appropriate weighting of partial bins in the numerator.

\subsection{Results}

\subsubsection{Full sample}
\label{sec:resFull}

\begin{figure}
\includegraphics[angle=-90,width=\linewidth]{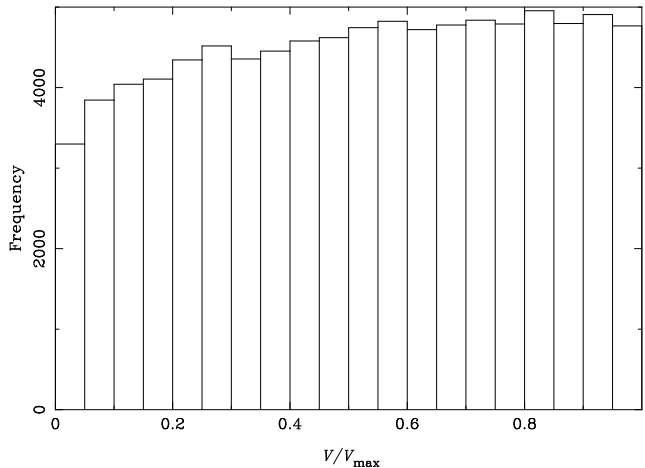}
\caption{Distribution of the $V/V_{\rm max}$ statistic for all main
sample galaxies in SDSS DR1.}
\label{fig:vvmax}
\end{figure}

\begin{figure}
\includegraphics[angle=-90,width=\linewidth]{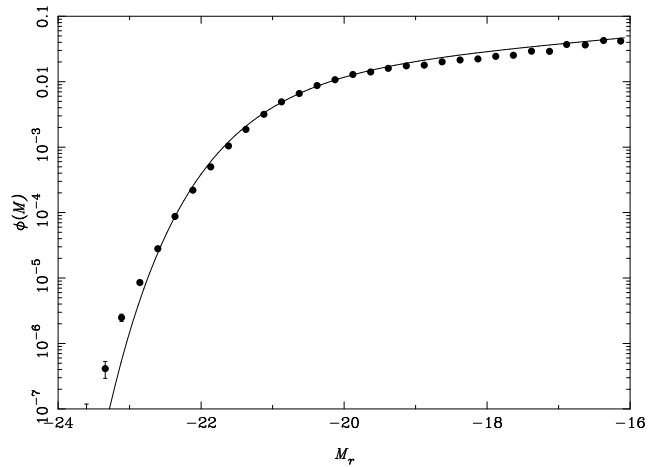}
\caption{Luminosity function for all main sample galaxies in SDSS DR1.
Symbols denote SWML estimates (the error bars are smaller than the symbol size
for most of the points) and the curve shows the best-fit Schechter function.
Note that no allowance for evolution has been made in this estimate.}
\label{fig:lf}
\end{figure}

We first estimate the luminosity function for all 90,275 galaxies in the 
SDSS DR1 with redshifts greater than 0.001 and with absolute magnitudes in the
$^{0.1}r$ band between $-24$ and $-16$.
The first indication of significant evolution in this sample comes from the
skewed distribution of the \citet{schmidt68}
$V/V_{\rm max}$ statistic (Figure~\ref{fig:vvmax}).
We find a mean $\langle V/V_{\rm max} \rangle = 0.523$, whereas one would
expect $\langle V/V_{\rm max} \rangle = 0.500 \pm 0.001$ for a homogeneous 
distribution of 90,275 galaxies.

The luminosity function of this full sample is shown in Figure~\ref{fig:lf}.
A Schechter function with parameters $\alpha = -1.23 \pm 0.02$,
$M^* = -20.63 \pm 0.02$ and $\phi^* = 0.0194 \pm 0.0010 h^3 \mbox{Mpc}^{-3}$
provides good agreement with the non-parametric SWML estimator except for
the bright end, $M_r < -22.5$, where we see in the SWML estimate a higher
density of galaxies than predicted by the Schechter function.
This bright-end excess can be explained by evolution of the LF:
the most luminous (rare) galaxies are likely to be seen at high redshift
and, as we shall see below, high-redshift galaxies have an enhanced luminosity
or density relative to low-redshift galaxies.
The Schechter function fit will be mostly constrained by galaxies at an
intermediate redshift ($\bar{z} \approx 0.1$).
Additionally, it is possible that some galaxies with extreme luminosities, 
$M_{^{0.1}r} \la -23$, may have incorrectly measured fluxes --- see below.

Note that any evolution in the luminosity function will render invalid
our assumption that the number density of galaxies can be separated into
a function of luminosity times a function of position.
In order to obtain an unbiased estimate of the LF, one needs to allow
for evolution.
\citet{blan2003L} have already shown that the $r$-band luminosity function 
of galaxies in the SDSS can be described by a Schechter function
whose characteristic luminosity $L^*$ brightens by $\approx 1.6$ magnitudes
per unit redshift.
Here we investigate the evolution of the galaxy LF in an alternative way, 
simply by subdividing the sample into slices in redshift.

\subsubsection{Redshift slices}

\begin{figure}
\includegraphics[angle=-90,width=\linewidth]{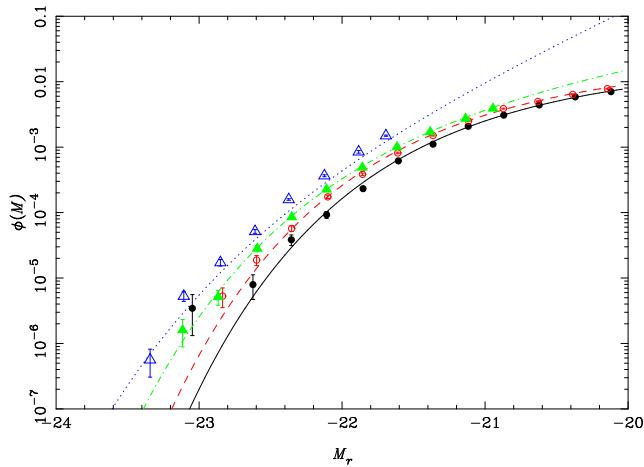}
\caption{Luminosity function for $z < 0.1$ (filled circles, continuous line),
$0.1 < z < 0.15$ (open circles, dashed line), 
$0.15 < z < 0.2$ (filled triangles, dot-dashed line),
$0.2 < z < 0.3$ (open triangles, dotted line).}
\label{fig:lfEvol}
\end{figure}

\begin{table}
 \caption{Number density $\bar{n}$ of galaxies in the range 
$-24 < M_{^{0.1}r} < -21.5$, in units of $10^{-4} h^3{\rm Mpc}^{-3}$.
For comparison, $\bar{n}_B$ shows galaxy density inferred from the
evolving Schechter LF of \citet{blan2003L}.
}
 \label{tab:dens}
 \begin{math}
 \begin{array}{crcc}
 \hline
 \bar{z} & N_{\rm gal} & \bar{n} & \bar{n}_B\\
 \hline
0.078 &  750 &  2.48 \pm 0.16 & 2.54 \pm 0.07\\
0.130 & 2477 &  3.64 \pm 0.15 & 3.33 \pm 0.09\\
0.178 & 5559 &  4.60 \pm 0.14 & 4.21 \pm 0.11\\
0.226 & 3325 &  7.36 \pm 0.19 & 5.26 \pm 0.14\\
  \hline
 \end{array}
 \end{math}
\end{table}

Figure~\ref{fig:lfEvol} plots the $r$-band luminosity function for galaxies
selected in four redshift slices: $0.001 < z < 0.1$, $0.1 < z < 0.15$,
$0.15 < z < 0.2$ and $0.2 < z < 0.3$.
The points with error bars show the SWML estimates, the lines show the
Schechter function fits.
Note that one cannot fairly compare the shapes of the Schechter fits
from this Figure, since the different redshift slices contain galaxies
covering a different range of absolute magnitudes.
In particular, the faint end slope of the highest redshift slice is
very poorly constrained, since there are no galaxies fainter than
$M_{^{0.1}r} \approx -21.5$ in this subsample.
If analysis of the intermediate $0.1$--$0.15$ redshift slice is limited to
$M_{^{0.1}r} < -21.5$, then the estimated faint-end slope changes from
$\alpha = -1.17 \pm 0.03$ to $-2.18 \pm 0.44$.
The Schechter fits in Figure~\ref{fig:lfEvol} are thus included for 
illustrative purposes only, and we deliberately do not quote the 
Schechter parameters.

The SWML data point for the lowest redshift slice at $M_{^{0.1}r} \approx -23$
lies significantly above the Schechter fit.
This is almost certainly due to errors in the photometry of the half-dozen
or so objects in this bin.
Of the ten apparently most luminous objects in the $z < 0.1$ slice
($M_{^{0.1}r} \la -22.6$), all have the \verb|EDGE| or \verb|COSMIC_RAY|
flag set.
Low redshift, luminous galaxies have much larger fluxes and apparent sizes
than most galaxies, and hence are more susceptible to their measured fluxes
being affected by cosmic ray hits or by lying near the edge of a CCD chip.

Despite the uncertainties in the {\em shape} of the LF
in the redshift slices, there is clear evolution in the {\em amplitude}
of the LF, in the sense of an increasing amplitude (vertical shift) and/or 
luminosity (horizontal shift) with redshift.
The estimated number densities of galaxies in the range 
$-24 < M_{^{0.1}r} < -21.5$,
obtained by applying (\ref{eqn:nbar}) to the non-parametric LF for each
redshift slice, are given in Table~\ref{tab:dens}.
For comparison, we also show the galaxy density inferred from the
evolving parametric LF of \citet{blan2003L}.
Their fit gives Schechter parameters $\alpha = -1.05$, $M^*_0 = -20.44$ and
$\phi^* = 0.0149 h^3{\rm Mpc}^{-3}$ at $z_0 = 0.1$ in the $^{0.1}r$ frame, 
with characteristic magnitude $M^*$ evolving as $M^* = M^*_0 - Q(z - z_0)$
with $Q = 1.62$.
We ignore the negligibly small number density evolution in their fit
and assume that the error in number density is dominated by the $\sim 3\%$
uncertainty in $\phi^*$.
Our density estimates agree within $\sim 2 \sigma$ for the three lower
redshift slices, but the density estimate for the highest redshift
slice is $\sim 9 \sigma$ larger than that inferred from the
Blanton et al. model.
The results presented here, and those of \citet{blan2003L}, 
are entirely consistent, since the Blanton et al. parametric model 
assumed linear evolution of the characteristic magnitude $M^*$ with redshift.
Less than 6\% of the galaxies in the full ($-24 < M_{^{0.1}r} < -16$)
sample are at $z > 0.2$, and so the Blanton et al. analysis would have
been insensitive to rapid evolution at these redshifts.

In the following section, we investigate evolution in the radial density
of galaxies in narrower redshift bins.
Note that the density estimator we use assumes a non-evolving
luminosity function, so any effect found can equally well be
ascribed to number or luminosity evolution.

\section{Radial density}  \label{sec:dens}

\subsection{Estimator}

Just as the maximum-likelihood estimate of the luminosity function $\phi(L)$ 
is independent of
inhomogeneities in the galaxy distribution, one can also estimate the radial
density $\rho(z)$ independently of the assumed luminosity function by 
maximizing the likelihood 
\begin{equation}
 {\cal L} = \prod_i \rho(z_i) \left/
            \int_{z_{\rm min}(L_i)}^{z_{\rm max}(L_i)} \rho(z) dV\right.,
  \label{eqn:rholike}
\end{equation}
where $z_{\rm min}(L_i)$ and $z_{\rm max}(L_i)$ are the limiting redshifts at
which a galaxy of luminosity $L_i$ would still be included in the survey. 
We fit
$\rho(z)$ by an arbitrary step function, using a variant of the SWML estimator
\citep{saunders90}. 
As in the maximum-likelihood estimate of
$\phi(L)$, overall normalization is lost and so we have applied the constraint 
\begin{equation}
 \int \rho(z) S(z) w(z) dV \left/ \int S(z) w(z) dV \right. = 1,
\end{equation}
where $S(z)$ is the selection function (\ref{eqn:selfn}) and
$w(z)$ is the weighting function (\ref{eqn:weight}).
This constraint is also used for the error estimates (cf. EEP).

\begin{figure}
\includegraphics[angle=-90,width=\linewidth]{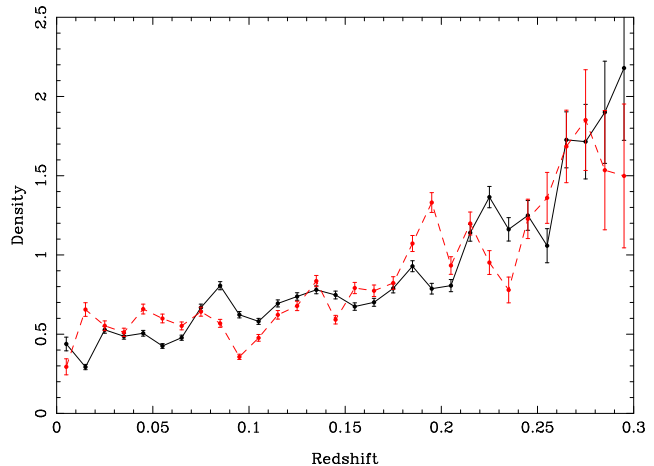}
\caption{Normalized comoving radial density plotted against redshift
for the Northern (continuous line) and Southern (dashed line) 
Galactic hemispheres.
Note that the radial density is normalized independently for the two
hemispheres.}
\label{fig:mldenNS}
\end{figure}

\begin{figure}
\includegraphics[angle=-90,width=\linewidth]{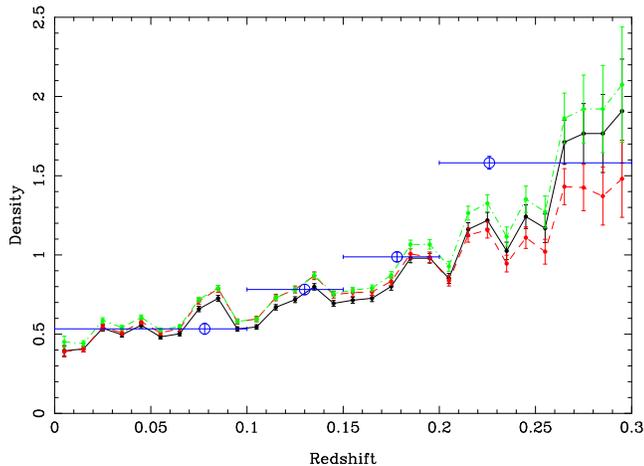}
\caption{Normalized comoving radial density plotted against redshift
for all galaxies assuming the same cosmology and $K$-correction scheme as 
\protect{Figure~\ref{fig:mldenNS}} (continuous line) and assuming a
$\Lambda=0$ cosmology (dashed line) and $K$-correcting to the $^{0.0}r$ band
(dot-dashed line).
The open symbols show normalized densities in the four redshift slices used
in the luminosity function analysis \protect{(Table~\ref{tab:dens})}.}
\label{fig:mldenTest}
\end{figure}

\subsection{Results}

Our estimate of $\rho(z)$ is plotted separately for the Northern and Southern
Galactic hemispheres in Figure~\ref{fig:mldenNS}.
We see a gentle increase in $\rho(z)$ out to $z \approx 0.2$, with a 
steeper increase to $z \approx 0.3$.
This estimate of radial density is independent of the galaxy luminosity
function only insofar as there is no correlation between redshift
and luminosity. 
The observed increase in radial density could thus either reflect an increase
in the number density of galaxies at higher redshift (whether due to number
density evolution or the existence of a large local ``hole''), 
and/or an increase in luminosity with redshift.
If there is a large local underdensity, it is extremely 
unlikely that we happen to lie exactly at the centre of it, and so
the consistent radial dependence of density in the Northern and 
Southern hemispheres provides strong evidence against the local hole 
hypothesis.

We have also checked the dependence of the estimated radial density on
the assumed cosmological model and our method of applying $K$-corrections.
Figure~\ref{fig:mldenTest} shows estimated radial density for the full sample
(north plus south) for the same assumed cosmology and $K$-correction as
Figure~\ref{fig:mldenNS}, along with the estimate separately assuming a
flat, $\Lambda=0$ cosmology, and $K$-correcting the $r$-band galaxy magnitudes
to redshift zero instead of redshift $z=0.1$.
The accelerated expansion in the $\Lambda$-dominated cosmology gives
rise to a larger density at high redshift compared with the $\Lambda=0$ 
cosmology.
Even though the increase in radial density with redshift is rather less if one
assumes a $\Lambda=0$ cosmology, particularly for $z \ga 0.2$, the increase
in density is still significant.
The estimated radial density is entirely consistent whether one $K$-corrects
galaxy magnitudes to the $^{0.0}r$ or $^{0.1}r$ bands, suggesting that
any errors in the $K$-correction have a negligible effect on the estimated
densities.

\begin{table}
 \caption{Results of a $\chi^2$ test for uniform density within two 
redshift ranges (low-$z$: 0--0.15 and high-$z$: 0.15--0.3)
for the DR1 sample and for clustered and random simulated catalogues.
In each case there are 14 degrees of freedom.
}
 \label{tab:chi2}
 \begin{tabular}{lrr}
 \hline
 Sample & low-$z$ & high-$z$\\
 \hline
 DR1 & 594 & 1730\\
 Clustered simulation & 400 & 32\\
 Random simulation & 14 & 11\\
  \hline
 \end{tabular}
\end{table}

We have tested the significance of the radial density evolution over the
redshift ranges 0--0.15 and 0.15--0.3.
In the lower redshift range, evolution is small but so are the error bars.
In the higher redshift range, where evolution is more pronounced, the
error bars are much larger.
For each redshift range, we calculate the mean radial density and
perform a $\chi^2$ test of the null hypothesis that all bins
have this mean density.
We apply the same test to two non-evolving, simulated catalogues.
The first consists of a random distribution of galaxies with comparable
galaxy numbers and LF to DR1.
The second consists of a \citet{sp78} hierarchical simulation with comparable 
galaxy numbers, LF, clustering and sky coverage to DR1.
The resulting $\chi^2$ statistics are given in Table~\ref{tab:chi2};
in each case there are 14 degrees of freedom (15 data points minus one
free parameter: the mean density).
\begin{description}
\item For the random simulation, the $\chi^2$ values are consistent with the
uniform density null hypothesis.
\item For the clustered simulation, the $\chi^2$ values are inconsistent with
uniform density, particularly in the low-$z$ range.
This is to be expected given the presence of 
large scale structure in the simulated galaxy distribution.
The biggest contribution to $\chi^2$ comes from the first three radial bins
($z < 0.03$) where density fluctuations are particularly pronounced
since clustering is not smeared out by peculiar velocities in these 
simulations.
\item The DR1 sample yields $\chi^2$ values marginally larger than the clustered
simulation in the low-$z$ range, and significantly larger than the clustered
simulation in the high-$z$ range.
Deviation from a non-evolving $\rho(z)$ is thus clearly very significant 
in the redshift range 0.15--0.30
but only marginally significant in the redshift range 0--0.15.
\end{description}

\subsection{Comparison with LF}

The open symbols in Figure \ref{fig:mldenTest} 
show the densities of luminous galaxies inferred when normalizing
the luminosity function in four redshift slices (Table~\ref{tab:dens}),
rescaled to have unit mean density.
The horizontal bars attached to these symbols denote the redshift range
(the symbol is centred on the mean redshift in each bin, 
rather than at the centre of each bin) and the vertical bars denote the
estimated error in density.
The agreement between these two completely independent methods of
estimating radial density is striking: the apparent increase in estimated
radial density with redshift can be explained entirely by evolution of
the galaxy luminosity function, with no need for a local underdensity.

\section{Galaxy Number Counts} \label{sec:counts}

\begin{figure}
\includegraphics[angle=-90,width=\linewidth]{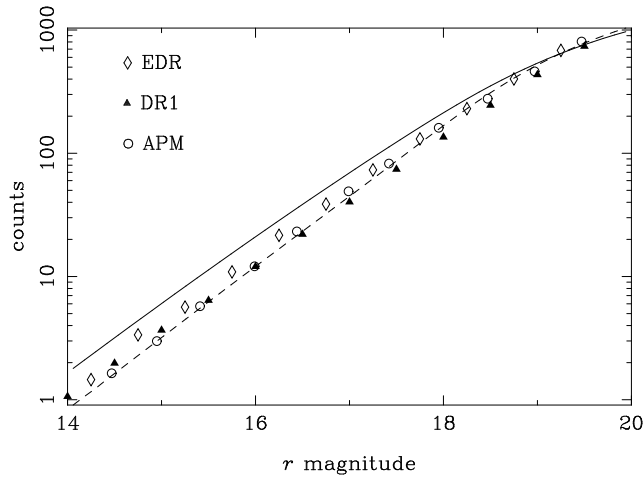}
\caption{Galaxy counts (per square degree, per unit magnitude)
as a function of $r$-band magnitude.
The continuous line shows predicted counts assuming a non-evolving 
luminosity function, the dashed
line assumes a radial density variation due to evolution as estimated
in Section~\ref{sec:dens}.
Observed counts are shown by diamonds (EDR), triangles (DR1)
and circles \citep[APM]{mselp90}.
For the APM counts, we have applied the very rough correction from 
$b_J$ to $r$-band magnitudes $r \approx b_J - 1.3$.}
\label{fig:counts}
\end{figure}

In this section we explore the effect of galaxy evolution, as reflected
in our estimate of radial density, on counts of galaxies as a function
of apparent magnitude.
In particular, galaxy number counts in the APM Galaxy Survey \citep{mselp90}
are steeper than one would expect unless there is significant evolution
of galaxies at low redshift.
Can the evolution seen here at $z < 0.3$ explain the steep APM counts?

The predicted number counts per unit magnitude are given by 
\begin{equation}
 n(m) = \int_0^\infty \phi[L(m,z)] \rho(z) \frac{dV}{dz}dz,
 \label{eqn:counts}
\end{equation}
where $L(m,z)$ is the luminosity of a galaxy at redshift $z$ and
with apparent magnitude $m$, $\rho(z)$ is the radial density
and $dV$ is the comoving volume element at redshift $z$.
These predicted counts are plotted in Figure~\ref{fig:counts} for
the non-evolving LF of Section \ref{sec:resFull}
with $\rho(z) \equiv 1$ (continuous line) and for
the same LF but with $\rho(z)$ as estimated for the full sample
in Section~\ref{sec:dens} (dashed line).
We make the conservative assumption when evaluating (\ref{eqn:counts})
that density does not evolve further beyond a redshift of 0.3, 
so that $\rho(z) \equiv 2.0$ for $z > 0.3$.

For comparison, the open circles show number counts measured in the $b_J$
passband from the APM Galaxy Survey \citep{mselp90}, where we have made a
very rough correction to the $r$-passband using $r \approx b_J - 1.3$.
The evolving model derived from SDSS data is in remarkable agreement 
with the APM counts considering the different passbands used and
areas of sky observed.
If the steep APM counts are due to a local underdensity, then a very similar
underdensity exists in SDSS galaxy counts which come predominantly
from the Northern sky.

Observed galaxy number counts from the SDSS Early Data Release
\citep{yasuda2001} are shown as diamonds (we have summed the Yasuda et al.
counts in the Northern and Southern equatorial stripes).
These counts are significantly shallower than the APM counts and are
consistent in shape with no galaxy evolution.
However, these counts come from an area of only 440 square degrees 
and so are susceptible to fluctuations from large scale structure.

We have therefore also estimated galaxy number counts from the 2099
square degrees of SDSS DR1.
Galaxies are selected from the DR1 database according to the following 
criteria:
\begin{enumerate}
\item None of the {\tt SATURATED}, {\tt BLENDED}, {\tt BRIGHT} or {\tt EDGE}
bits be set in the $r$-band {\tt photoFlags}.
\item The object be classified as a galaxy in at least two of the
$g$, $r$ and $i$ bands.
\end{enumerate}
These counts are shown as triangles in Figure~\ref{fig:counts}.
While not quite as steep as the APM counts, they still lie closer
to the evolving LF model than to the non-evolving model.

Rapid evolution of the luminosity and/or density of the galaxy population 
at redshifts
$z < 0.3$ thus provides a natural explanation for the observed radial
density of SDSS galaxies and steep number counts of APM and DR1 galaxies.

\section{conclusions} \label{sec:concs}

We have presented evidence for significant evolution in the luminosity function
of $r$-band selected galaxies in the SDSS DR1 at redshifts $z < 0.3$.
This evolution gives rise to a factor $\sim 3$ increase in inferred
galaxy density between redshifts 0 and 0.3, in agreement with earlier
findings from \citet{eales1993} and \citet{blan2003L}.
Folding this evolution into predictions for galaxy number counts 
gives a remarkably good match to the slope of number counts observed
in the APM Galaxy Survey and in DR1 itself.

The aim of this short paper has simply been to demonstrate that the
luminosity function of galaxies has evolved significantly at recent
epochs, since $z < 0.3$, and that such evolution is sufficient to
explain the steepness of the observed number-magnitude counts of APM
galaxies, without the need to invoke a local underdensity or a
magnitude scale error\footnote{The possibility of a significant scale
error in the APM magnitudes has recently been ruled out using CCD
photometry \citep{love2003}.}.

Note that these results do not preclude the existence of significant
density fluctuations in the local Universe on very large scales.
Indeed, \citet{fbfms2003} have recently used galaxies in the 2 Micron All Sky
Survey to indicate the presence of a region in the Southern Galactic
hemisphere $\sim 200 h^{-1} {\rm Mpc}$ in extent 
with a mean underdensity $\sim 30\%$.
This underdensity alone, however, is insufficient to explain the
steepness of the APM counts.
We have shown that evolution of the galaxy LF, possibly in combination
with large density fluctuations, can explain the steep counts.

Here we have not attempted to investigate the type of evolution that
is occurring: the possibilities include any combination of luminosity
evolution, density evolution and change in shape of the galaxy LF.
Future work will use SDSS data to study galaxy evolution in detail.
We plan to investigate luminosity evolution in different passbands and 
for different galaxy types, and to perform a detailed investigation of 
the spectral evolution of galaxies with redshift.

\section*{Acknowledgments}

It is a pleasure to thank Mike Blanton for help with \verb|kcorrect|,
Jim Gray, Robert Lupton and Ani Thakar for advice on the galaxy number counts
database query and Mike Blanton, Yeong-Shang Loh and Bob Nichol for their 
comments on an earlier draft of this paper.

Funding for the creation and distribution of the SDSS Archive has been
provided by the Alfred P. Sloan Foundation, the Participating
Institutions, the National Aeronautics and Space Administration, the
National Science Foundation, the U.S. Department of Energy, the
Japanese Monbukagakusho, and the Max Planck Society. The SDSS Web site
is \verb|http://www.sdss.org/|.

The SDSS is managed by the Astrophysical Research Consortium (ARC) for
the Participating Institutions. The Participating Institutions are The
University of Chicago, Fermilab, the Institute for Advanced Study, the
Japan Participation Group, The Johns Hopkins University, Los Alamos
National Laboratory, the Max-Planck-Institute for Astronomy (MPIA),
the Max-Planck-Institute for Astrophysics (MPA), New Mexico State
University, University of Pittsburgh, Princeton University, the United
States Naval Observatory, and the University of Washington.

{}

\end{document}